\documentclass[final,1p,times]{elsarticle}
\usepackage{epsfig}
\usepackage{amssymb}
\usepackage{graphics}
\usepackage{amsthm}
\journal{Physica A}

\begin{document}
\begin{frontmatter}
\title{Low energy modes and Debye behavior in a colloidal crystal}

\author[1]{Antina Ghosh}, \author[2]{Romain Mari},
\author[1]{V. K. Chikkkadi},\author[1]{Peter Schall},
\author[3]{A. C Maggs},\author[1,4]{Daniel Bonn}

\address[1]{Van der Waals Zeeman Institute, Univ.\ of Amsterdam, Valckenierstraat 65, 1018 XE
Amsterdam, the Netherlands.}\texttt{}\\
\address[2] {PMMH UMR7636, CNRS, ESPCI, 10 rue Vauquelin, 75005, Paris, France.}
\address[3]{Physicochimie th\'eorique, Gulliver, CNRS-ESPCI, 10 rue
Vauquelin, 75005, Paris, France.}\\
\address[4]{LPS de l'ENS, CNRS UMR 8550, 24 Rue Lhomond, 75005 Paris,
France.}

\begin{abstract}
We study the vibrational spectrum and the low energy modes of a three dimensional
colloidal crystal using confocal microscopy. This is done in a two-dimensional cut through a
three-dimensional crystal. We find that the observed density of
states is incompatible with the standard Debye form in either two
or three dimensions. These results are confirmed by numerical
simulations. We show that an effective theory for the projections
of the modes onto the two-dimensional cut describes the
experimental and simulation data in a satisfactory way.
\end{abstract}

\end{frontmatter}

\section{Introduction}
\label{sec1} Colloidal particles dispersed in a solvent form fluid
and solid phases under appropriate conditions \cite{VanMegen,
Torquato}. The ordered phase is interesting due to their
long-range order combined with very soft mechanical properties
making thermal fluctuations very important. For the colloidal
systems considered here, the shear modulus is only a few $Pa$,
whereas crystalline solids typically have moduli in the $GPa$
range\cite{Clark, Allard}. Comparison of the properties of
colloidal systems with harder molecular solids is a matter of
ongoing research \cite{Keim,Zahn,Chaikin}. Earlier studies have
largely focused on the the phonon dispersion behavior measured by
means of video microscopy \cite{Keim} or light scattering
\cite{Chaikin2, PENCIU}. However, to our knowledge there is no
experiment that directly verifies the Debye scaling in the
measured density of states of the vibrational modes. Even for
molecular crystals, the this has proven rather hard to measure
directly; mostly neutron scattering experiments have provided some
but not much data that  agree with the expected $\omega^2$
behavior \cite{Ghatak}. Perhaps for this reason the temperature
dependence of the specific heat is usually taken as the hallmark
for the Debye behavior of normal crystalline solids. The question
we ask in this paper is how the spectrum measured in a colloidal
crystal compares with the predicted Debye behavior, $D(\omega)
\sim
\omega^{d-1}$.\\
To study the energy spectrum, we calculate the normal modes of a
hard sphere colloidal crystal from the correlations in particle
displacements. The colloidal particles used are small enough so
that they perform small `vibrations' in response to the thermal
excitations of the surroundings. We use confocal microscopy to
observe and record these particle motions in a two dimensional
field of view. The eigenvalues of the dynamical matrix and normal
modes are then obtained from the spatial correlations \cite{Keim,
ghosh1}of the measured displacements. The present experimental
results are compared and complemented with a simplified continuum
theory as well as Monte Carlo simulation of hard sphere crystals.
The main conclusion is that the spectrum of slice of a three
dimensional system has a anomalous density of states which varies
as $D(\omega)\sim \omega^3$, which contrasts the assumption made
in \cite{science} that a $2d$ cut should exhibit the expecred $3d$
Debye behavior. However our experimental results do agree with
simulations presented below, and can in addition be explained by a
simplified continuum theory.

\subsection{Experiments}

Hard sphere colloidal systems undergo phase changes with the
volume fraction $\phi$ as control parameter. There is no
liquid-gas transition, but fluid-solid coexistence \cite{VanMegen,
Torquato} is observed from the freezing transition point
$\phi_{f}=0.494$ until the melting point $\phi_{m}=0.545$. A
stable crystalline phase exists above $0.545$ until the closed
packed density at $\phi=0.74$. The colloids we use are charge
stabilized PMMA particles with a diameter of $a=1.3 \mu m$ with a
very small size polydispersity of about $2\%$. The particles are
dyed with rhodamine and are suspended in a CHB (cyclohexyl
bromide) / decalin mixture which closely matches both the density
and the index of refraction of the particles. We further add
organic salt TBAB (tetrabutylammoniumbromide)
to screen any possible residual charges.\\
\begin{figure}[h]
  \begin{center}
    \includegraphics[width=5.0cm, height=5.0cm]{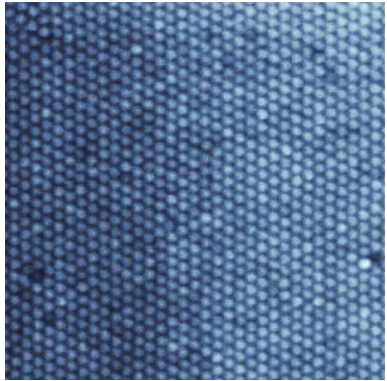}
    \caption{Snapshot of a two dimensional section in a field of view
      of $60\mu m \times 60 \mu m$ with a hexagonal symmetry.}
    \label{2Dimage}
  \end{center}
\end{figure}
The crystal was grown in a sample cell made of parallel plates with
a confinement of approximately $\sim 1 mm $ along the vertical direction. The
volume fraction of the present colloidal crystal is about $\phi
\approx 0.57$. Using confocal microscopy we acquire images at a
speed of $25$ frames per second of a two dimensional section of
about 60 $\mu m$ x 60 $\mu m$ of the larger three dimensional
crystal. The 2D slice was taken at a distance of $25-30$ micron
away from the coverslip, deep enough to avoid the effects of confinement.
The entire crystal is polycrystalline, but we take our
data from a region of the crystal that as far as we can see is
perfectly crystalline and contains no defects. The particle
positions are identified and tracked for a period of about $120$
seconds using standard particle-tracking software; this results in
a total of $3000$ frames. Fig. \ref{2Dimage} shows a typical
snapshot of a two dimensional section of the measured crystal.
\subsection{Density of states}
We obtain an ensemble of projected particle positions
${r}_i=\{x_{i}, y_{i}\}$ from the above measurements. The
displacement components from the mean position for the $i^{th}$ particle
is given by,
\begin{equation}
  u_{\mu}(i) = r_{\mu i} - \langle r_{\mu i} \rangle.
\end{equation}
 $$
  u_{\nu}(i) = r_{\nu i} - \langle r_{\nu i} \rangle.
 $$
where $\mu, \nu = x, y$  and ``$\langle \rangle$'' indicates an ensemble average. The
$k^{th}$ Fourier component of the above displacements is given by,
\begin{equation}
 u_{\mu}(k) = \sum_{i=1}^{N} u_{\mu}(\vec{r_{i}})
 exp(-j\vec{k}.\vec{r_{i}})
\label{Fcomp}
\end{equation}
$$
 u_{\nu}(k) = \sum_{i=1}^{N} u_{\nu}(\vec{r_{i}})
 exp(-j\vec{k}.\vec{r_{i}}),    k = \{k_{x}, k_{y}\}
$$
The summation above runs over all the $N = 1386$ colloidal particles
present in the system and the respective
$k$ values are chosen from the first Brillouin zone of the
experimental lattice.
\begin{figure}
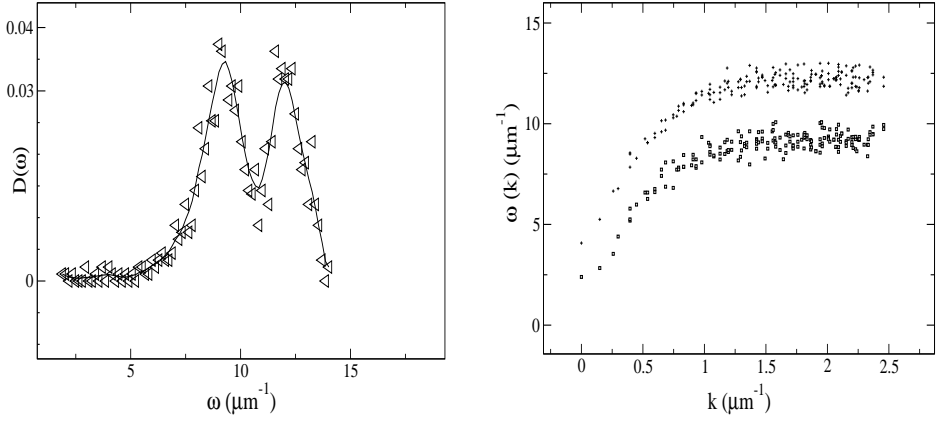

  \centering
  \begin{minipage}{6.0cm}
    \includegraphics[width=5.8cm,height=5.5cm]{D2-DOS.eps}
  \end{minipage}
  \quad
  \begin{minipage}{6.0cm}
    \includegraphics[width=5.8cm,height=5.5cm]{D2-dispers.eps}
  \end{minipage}
  \caption{Left: Density of states as obtained from the correlations,
    eq.~\ref{DOSEq1}. Right: The dispersion curves of transverse and
    longitudinal vibrations. The frequencies (vertical axis) near the
    zone boundary where $d\omega/dk \approx 0 $ coincide with the
    positions of the singularities (peaks) in the density of states on
    top.}
 \label{FDOS}
\end{figure}
 Now, the potential energy of a harmonic crystal can be written as,
\begin{equation}
  U = 1/2 \sum_{k}(u^{*}_{\mu}(k)K_{\mu \nu}(k)u_{\nu}(k))
\end{equation}
where $K_{\mu \nu}(k)$ \cite{Keim}is the $2 \times 2$ dynamical
matrix in Fourier space.
 From equipartition each of the above quadratic terms
$u^{*}_{\mu}(k)K_{\mu \nu}(k)u_{\nu}(k)/2$ contains an
energy of $k_BT/2$. Therefore we can write,
\begin{equation}
  K^{-1}_{\mu \nu}(k) =\left \langle
     {u}^{*}_{\mu}(k){u}_{\nu}(k)\right\rangle /k_BT .
  \label{DOSEq1}
\end{equation}
In the present study, we take our two-dimensional data and analyse
it according to the above scheme. The mode frequencies of the
system are related to the
eigenvalues $\lambda$ of the matrix $K_{\mu \nu}^{-1}(k)$ as,
\begin{equation}
  \omega(k)  = \sqrt{1/\lambda(k)}.
  \label{freq}
\end{equation}

Now, there are $N$ of such $2 \times 2$ matrices ($K_{\mu
\nu}^{-1}$) corresponding to the number $N$ of $k$ values.
Separate diagonalization of each of these gives us $2N$ normal
mode frequencies.
 Each of the $N$ branches of
frequencies as a function of $k$ then leads to the two dispersion
curves shown in Fig. \ref{FDOS}. The density of states $D(\omega)$
as obtained from these frequencies is shown in the same figure. A
two peak structure is apparent in the density of states. Such
peaks are familiar from the theory of lattice dynamics and often
signal van Hove singularities \cite{Ghatak} occurring due to the
vanishing group velocities $\nabla_{k} \omega$ at certain wave
vectors. Indeed we see that the frequencies where $d\omega/dk
\approx 0$ (close the zone boundary) on the dispersion curves
coincide with the ones around which peaks appear in the density of
states.

One important additional remark is that if we make the reasonable
assumption that the system is in local equilibrium, the values of
the squared-displacements are independent of the dynamics, which
may be overdamped or purely ballistic, and even contain
hydrodynamic interactions. What depends on the nature of the
dynamics is the actual time-dependence of the displacements, but
not their statistical distribution. Our new method \cite{ghosh1,
ghosh2} directly analyzes the normal modes of the vibrations in
the cages from the displacement correlations. If the system were
harmonic, undamped and without hydrodynamic interactions, the DOS
obtained from e.g Fourier transforming the velocity
autocorrelation function and our method would coincide. However,
in the former method all the spatial in- formation is lost. The
strength of the present analysis using confocal measurements lies
in directly visualizing the collective modes at low spatial
frequencies and therefore gaining information about the nature of
such modes. But the price to pay for this is the exclusion of
knowledge of damping, anharmonicity and hydrodynamic interactions.

\subsection{Visualization of modes}
The next question we ask is what the low energy modes of the above
system look like ? To be able to visualize these modes we compute
the spatial correlation between the particles in real space via
the covariance matrix.
\begin{figure}[!h]
  \begin{center}
    \begin{minipage}{5cm}
      \includegraphics[width=5cm,height=5.0cm]{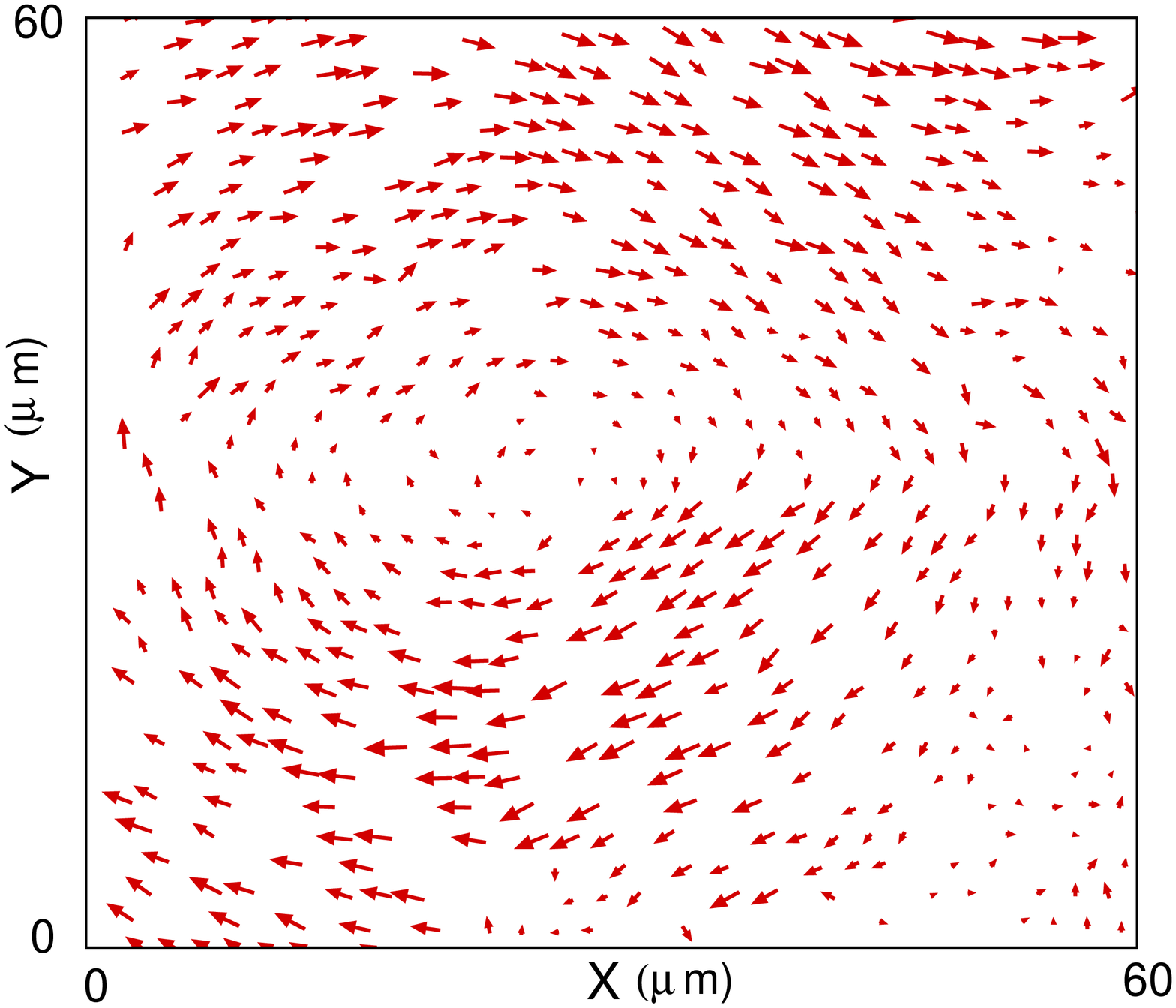}
      \center{a}
    \end{minipage}
    \begin{minipage}{5cm}
      \includegraphics[width=5cm,height=5.0cm]{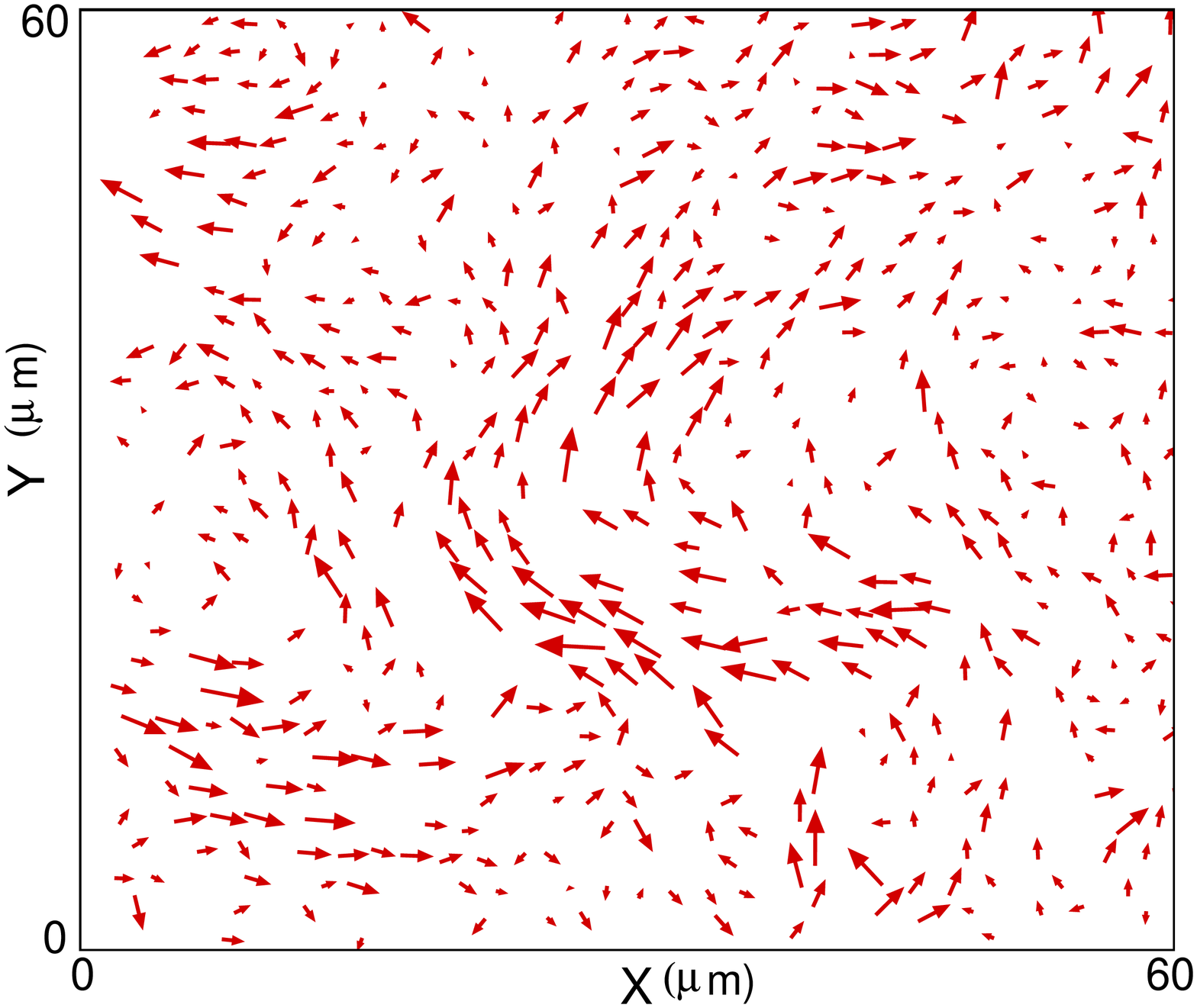}
      \center{b}
    \end{minipage}
    \begin{minipage}{5cm}
      \includegraphics[width=5.0cm,height=5.0cm]{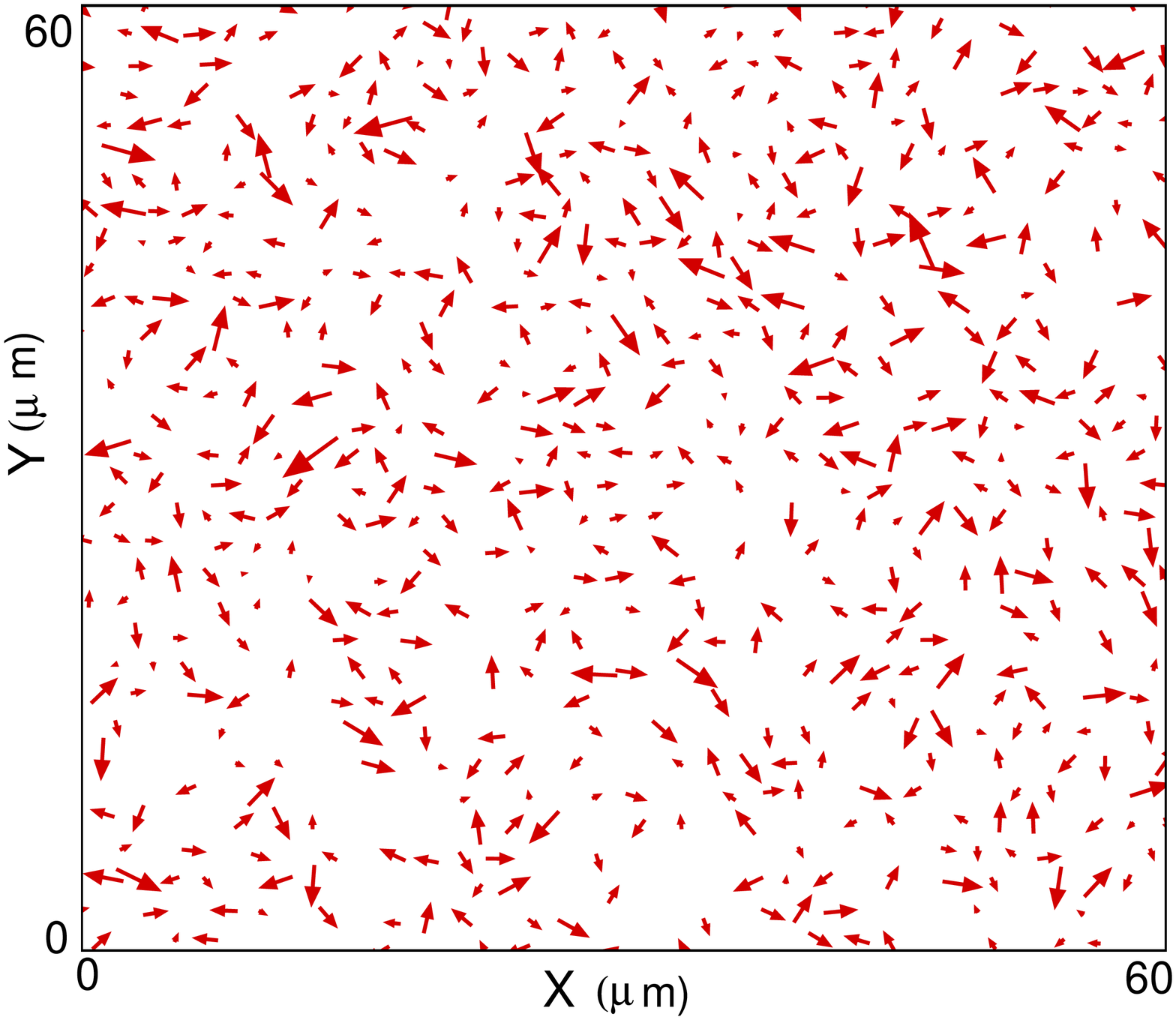}
         \center{c}
    \end{minipage}
    \caption{Tomography of the normal modes shown in a
    field of view of $60 \mu m \times 60\mu m$. The pictures correspond to increasing frequencies going from a) to
     b) to c).
     The very low
      frequency ones show the extended plane wave nature of the
      modes. This coherent character is lost and modes appears to be
      random as higher frequencies are approached.}
     \label{Eigen}
  \end{center}
\end{figure}
This matrix, which has been used recently in
studying normal mode properties of colloidal \cite{Keim, KeChen,
science, ghosh1, ghosh2} and granular systems \cite{Brito}, is
defined as:
\begin{equation}
  Q_{\mu \nu}(i,j) = \left \langle u_{\mu}(i)) u_{\nu}(j)\right \rangle .
  \label{SQ}
\end{equation}
This is a $2N \times 2N$ dimensional matrix for $N$ particles. Any
eigenvector $\{v_{l}\}$ of the above matrix represents a normal
mode at a single ``frequency'' $\omega_{l} =
\sqrt{1/\lambda_{l}}$.  In our colloidal system, it allows us to
\textit{compute} the normal modes, rather than supposing plane
waves as in eq.~(\ref{DOSEq1}). A few examples of the normal modes
are shown in Fig.~(\ref{Eigen}). The modes in the low-frequency
part of the spectrum show a clearly coherent motion extending over
large part of the field of view.  For higher frequencies the modes
appears to be rather random in nature.
\newpage
\subsection{Scaling of the low frequency density of states}
Having described the vibrational spectrum of the experimental
crystal we come back to one of the central questions of this
paper, which is what happens if one tries to study the matrix
eq.~(\ref{DOSEq1}) using only a subset of the particles of the
true three dimensional system, in particular if one reconstructs
the dynamical matrix from the measurements from a single plane
within the sample as we do. What is the nature of the effective
interactions found in the analysis and what is the density of
states found in the two-dimensional slice ?
\begin{figure}
\begin{center}
\begin{minipage}{6.0cm}
    \includegraphics[width=6.cm,height=5.5cm]{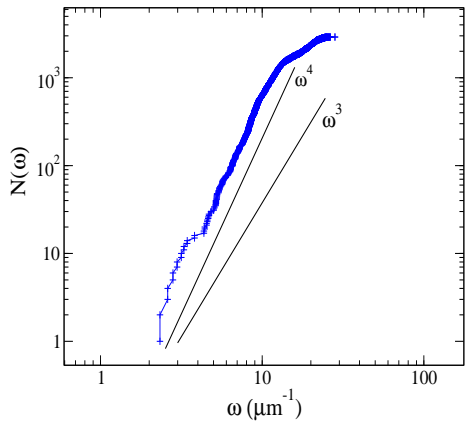}
   \end{minipage}
\caption{A power-law fit to the low frequency part of the
    cumulative density of states $N(\omega)$.  The measured exponent
    $(N(\omega)\sim \omega^{\alpha})$, is $\alpha \approx 3.93 \pm
    0.19$.}
\end{center}
\end{figure}
To answer this we look at how the measured cumulative density of
states of the colloidal crystal compares to the expected
integrated density of states $N(\omega)  = \int D(\omega) d\omega
\sim \omega^{3}$ in $d=3$ We find that the measured $N(\omega)$
scales approximately as $\omega^{4}$ rather than $\omega^{3}$. We
fit the integrated density of states to a power law with
adjustable exponent, $\alpha$.  Our data best fits to a value
close to $\alpha=3.9 +/- 0.19$, far from the expected value
$\alpha=3$ for a three dimensional elastic medium and even further
from the value $\alpha=2$ expected in two dimensions. The quoted
error is the one-sigma deviation of the Gaussian distribution of
the scatter of the data around the fit.  A standard error
calculation gives that the $99$ percent probability interval is
$3.93 +/- 0.416$, excluding the power $3$. Furthermore, we have
done a few different measurement series on the same sample, but at
different locations within the sample and size of the region
probed. Each of these measurements was analyzed separately to
ensure the robustness and consistency of the results. Taking data
from different regions in the sample, the exponent was found to
always lie in the interval $[3.7 -4.1]$, with a mean value of
$3.9$. This again confirms the value of the exponent and its error
quoted above.

The above scaling of the low frequency cumulative DOS is also
checked with the spectrum from the spatial covariance matrix that
contains the non-diagonal correlations(sec.\ref{methods}). This
yields an exponent of about $3.6 +/- 0.3$ which is reasonably
close to the scaling $(3.93 +/- 0.19)$. Thus, within the
experimental error, the scaling of the DOS from both the full
diagonalization and $K_{\mu \nu}^{-1}$ are the same, and different
from the usual Debye scaling. This new result will be further
discussed and explained below, using Monte Carlo simulations of
hard sphere crystals and arguments from elasticity theory.

\section{Theory and Simulation}

\begin{figure}
  \begin{center}
    \begin{minipage}{6cm}
      \includegraphics[width = 5.5cm, height = 4.5cm]{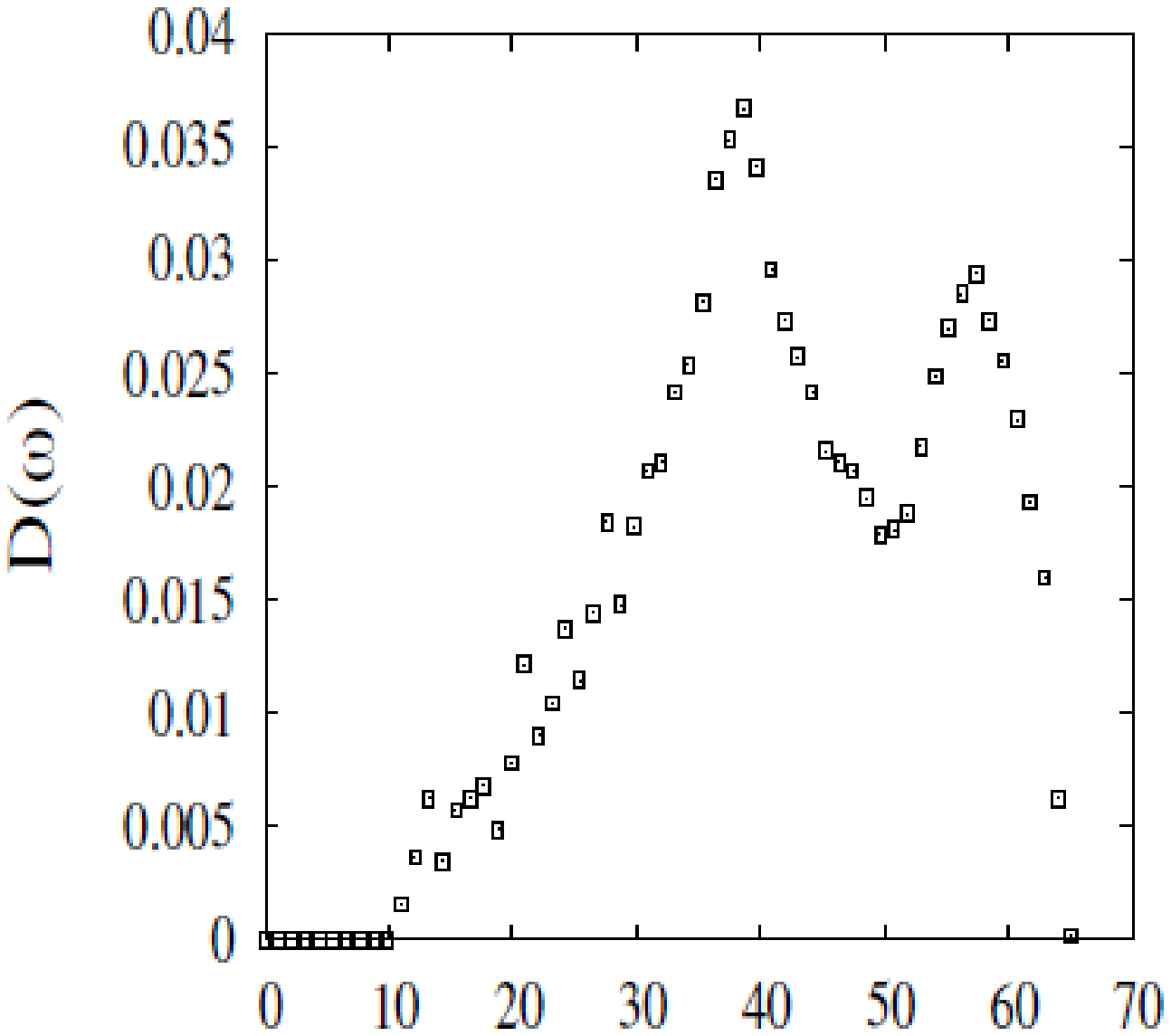}
    \end{minipage}
    \begin{minipage}{6cm}
      \includegraphics[width = 5.5cm, height = 4.8cm]{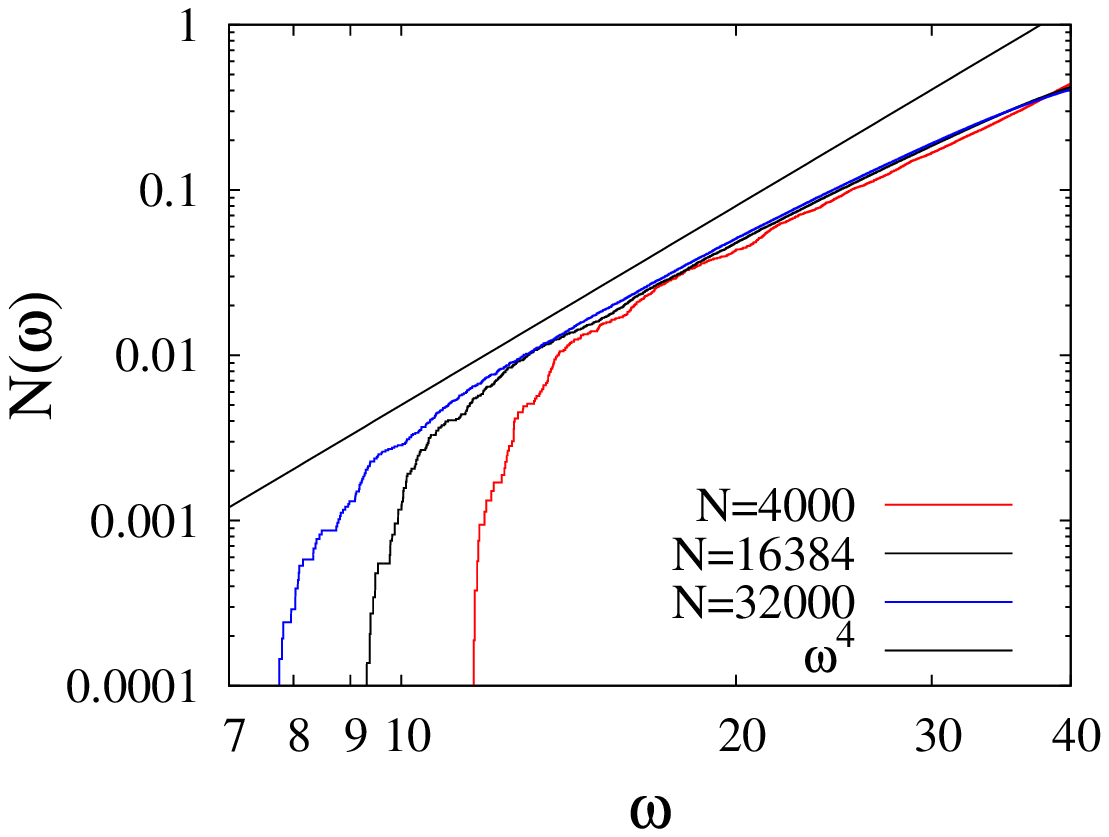}
    \end{minipage}
      \caption{Left: Density of states for a two dimensional section
of a 3D hard sphere crystal (with $N=4000$) from Monte-Carlo
simulation. Right: Low frequency part of the cumulative
DOS for two dimensional slices of the three dimensional system
for three system sizes, with logarithmic scale, showing a powerlaw
behavior compatible with the Debye behavior $N(\omega)\sim \omega^{4}$.
Note that the range over which the scaling holds increases with
the system size, indicating a convergence to the expected result.
Frequencies are in unit of the inverse particle diameter.}
    \label{sim-DOS}
  \end{center}
\end{figure}

We check our experimental findings using Monte-Carlo simulation of
a hard sphere crystal. We use a face-centered cubic crystal made
of 864 particles, with periodic boundary conditions.  We thermally
equilibrate the system over a sufficient number ($10^6$) of MC
steps.  We compute the covariance matrix (Eq. \ref{SQ}) to find
the eigenmodes of the system along runs of $10^5$ MC sweeps, out
of which we use $5\times 10^4$ snapshots for time averaging. We
obtain the matrix for both the whole three dimensional system and
two dimensional $(111)$ planes, in order to mimic the experimental
conditions (see also \cite{ghosh2} for a comparison between the
$3d$ simulations and a $2d$ cut). As the number of particles (and
thus modes) in a two dimensional plane is quite small, we average
over several planes. The density of states we obtain is shown in
Fig. \ref{sim-DOS}. For the low frequency part of the cumulative
DOS of the two dimensional slice we again find a cumulative
density of states which seems to converge to $N(\omega) \sim
\omega^4$ when we increase the system size, which is compatible
with the experimental data. Data with $N=32000$ can be fitted by
exponents between $3.7$ and $4.1$."

We now turn to a simple scalar theory which gives an indication as
to  how Debye theory must be modified when one observes two
dimensional cuts of a three  dimensional sample.

The elastic energy of a three dimensional crystal can be written
in terms of a symmetric shear tensor $u_{i,j}$ and three
independent elastic constants. At large length scales correlations
in displacement fluctuations decay as $1/r$,  but are also
characterized by a complicated tensorial structure coming from the
cubic anisotropy of the crystal.

The effect of the large distance decay can be found in a much
simpler  theory based on a scalar field $u$ rather than the vector
$u_i$. This scalar can be thought of as being, for instance,  the
amplitude of longitudinal fluctuation which couple to density
fluctuations. The advantage of such a description is an enormous
simplification in the tensorial algebra and a simple closed form
for the projected correlation function. It is possible to perform
a detailed tensorial calculation, which we will publish in the
future and which yields very similar results.

We therefore consider fluctuations of a scalar quantity $u$ with
an energy which is of the form
\begin{equation}
  U =  \frac {A}{2} \int (\nabla u)^{2} d^{3}\bf{r}
 \end{equation}
where $A$ is an elastic modulus. In the generalization to elastic
fluctuations one would consider an energy based on the symmetrized
strain tensor.  In Fourier space the energy has the form
\begin{equation}
  U =  \frac{A}{2} \sum_{k} k^2 |u_{k}^{2}| \label{en}
\end{equation}
we notice the usual scaling of the elastic energy in $k^2$.

In an underdamped system with kinetic energy $\rho \dot u^2/2$
this gives rise to the dispersion relation $\omega^2 = c^2 k^2$.
One thus expects a density of states
\begin{equation}
  {\rm d} N \sim k^2 \, {\rm d}k \sim \omega^2 {\rm d} \omega
\end{equation}
It is this scaling of the density of states in $\omega^2$ that is
known from the theory of Debye.

In a three dimensional sample thermal fluctuations excite the system,
so that equipartition and eq.~(\ref{en}) implies
\begin{equation}
  \langle |u_{k}|^2\rangle = \frac{k_BT}{ A k^{2}}
\end{equation}
This gives a decay of correlations in real space which is given by the
inverse Fourier transform of $1/k^2$. We can find the result
immediately by reference to electrostatics:
\begin{equation}
 \langle u(i) u(j) \rangle = \frac{1}{4\pi A |r_{i} -r_{j} | }
\end{equation}
a Coulomb like decay of correlations.

Now take a two dimensional slice of the system. Within this slice the
correlations are still decaying as $1/r$. We wish to describe what we
see, however, in terms of a purely two-dimensional theory, so we
perform a two dimensional Fourier transform to find the effective
stiffness. Thus
\begin{equation}
  \langle |u_{k}|^2\rangle_2 =  \int \frac{1}{Ar}\,
  e^{i{k}_2 \cdot {r}} \,
  d^2\, {r}\,
  = \frac{1}{2A|k_{2}|}
\end{equation}
where we use the subscript $2$ to indicate that we are working
with the two-dimensional projected objects.  The result is rather
interesting: rather than correlations in three dimensional being
described by a decay in $1/k^2$ we find a slower decay: $1/|k_{2}|$.

Now that we have the scaling form of the correlations we can work
backwards and deduce the effective elastic theory in two
dimensions
\begin{equation}
  U_{2} =  \frac{A}{2} \sum_{k_2} 2|k_{2}| |u_{k_{2}}|_{2}^{2}
\end{equation}
Thus the elastic behavior in real space corresponds to fractional
derivatives of the field $u$ leading to long-ranged effective
interactions in the projected system.

We now calculate the ``propagative'' eigenvalue by defining
\begin{equation}
  \omega_{2}^{2} = 2A |k_{2}|
\end{equation}
which is the analogy of
\begin{math}
  \omega^{2} = c^{2} k^{2}
\end{math}
that we use in three dimensions. We note that the dispersion law
is very different from that of usual elastic problems.  The
density of states of this two dimensional matrix are just
\begin{equation}
  {\rm d} N_{2} \sim k_{2} {\rm d} k_{2} \sim \frac{\omega_{2}^{3}}{A^{2}} {\rm d}\omega_{2}
\end{equation}
The density of states is thus $D(\omega)\sim \omega_{2}^{3}$ and the
integrated density of the states  $N(\omega) \sim \omega_{2}^{4}$, in
good agreement with what was found in the experiment and in the
simulations.
\subsection{Comparision of methods}
\label{methods}
In this section we briefly discuss and compare
the two methods adopted in the present study to obtain
the spectrum namely the $ 2 \times 2 $ dynamical matrix
$K_{\mu \nu}^{-1}(k)$ and the spatial covariance
matrix described in Eq. \ref{SQ}. To establish a correspondence we compute the full
$2N \times 2N$ matrix in k-space,
\begin{equation}
\tilde{Q}_{\mu \nu}(k,k') = \langle
u^{*}_{\mu}(k)u_{\nu}(k') \rangle.
\label{FQ}
\end{equation}
\begin{figure}[!h]
  \begin{center}
    \begin{minipage}{5cm}
    \includegraphics[width=5.0cm,height=5.0cm]{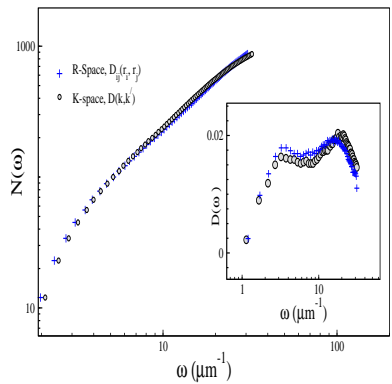}
     \end{minipage}
    \caption{Normal mode spectrum: Cumulative Density of
      states and DOS for the present crystal as obtained from the covariance
      matrix $Q_{\mu \nu}(r_{i},r_{j})$ in real space and
      its equivalent $\bar{Q}(k,k')$ in Fourier space. For
      a perfect crystal these two methods should give identical
      results. In the case of imperfections the modes $|k\rangle$ and
      $| k'\rangle$ are no longer orthonormal leading to small differences
      in the spectra. }
    \label{Dw-compr}
  \end{center}
\end{figure}
using the same Fourier components of displacements as described
earlier (Eq. \ref{Fcomp}), which includes the non-orthogonal
correlations as well. Let us first discuss the results obtained
from the full diagonalization of both $\tilde{Q}$ matrices (Eqs.
\ref{SQ}  and \ref{FQ}). Fig. \ref{Dw-compr} compares the spectrum
- DOS and cumulative DOS for both the methods: they nearly
coincide. This confirms the equivalence of the spatial matrix
$Q_{\mu \nu}(i,j)$ with $\tilde{Q}_{\mu \nu}(k,k')$ and supports
the fact that the data are not affected by the choice of the
boundary conditions or imperfections of the crystal. Now, the
diagonal elements of $\tilde{Q}_{\mu \nu}(k,k')$ correspond to
$K_{\mu\nu}^{-1}$ but the non-diagonal elements in $(k,k')$ encode
information about the heterogeneity and imperfections of the
sample. Thus, with negligible imperfections in the crystal and a
sufficiently long averaging time, the spectrum from all three
methods should coincide with each other.
\subsection{Conclusions}
We have studied the dynamics of a hard sphere colloidal crystal at
a volume fraction $\phi \sim 0.57$, a volume fraction slightly
above the melting transition, using confocal microscopy. The
density of states and normal modes were obtained from measured
particle displacements. Hard sphere systems are usually weakly
connected and the interaction potential is strongly anharmonic;
however the present observations shows that the lowest frequency
modes are extended plane waves-like as can be expected for a
harmonic solid. In addition we have shown that the density of
states can be understood using continuum elasticity theory.

The effective exponent for the frequency-dependence of the density
of states was measured in the low energy regime and is
inconsistent with the expected Debye behavior in $D(\omega)\sim
\omega^{d-1}$ for both $d=2$ and $d=3$. We found that the data can
be explained by a theory with an unusual energy dispersion
relation in $|k_{2}|$, which gives $D(\omega) \sim \omega^3$. This
expression agrees with both the experiments and the simulations.
It is interesting to note that the same energy function was found
in \cite{jf} where the spreading of a droplet was expressed as the
effective dynamics of a contact line. Again we are in the presence
of a physical system projected to lower dimensions.

We appreciated discussions with  Jorge Kurchan and Gerard Wegdam.
The present project is supported by FOM.

\end{document}